\documentclass[aps,floats]{revtex4}
\usepackage{amsmath,amssymb}
\usepackage{graphicx,epsfig}

\begin{document}
\bibliographystyle {plain}

\def\oppropto{\mathop{\propto}} 
\def\opsimeq{\mathop{\simeq}}
\def\opoverderline{\mathop{\overline}}
\def\operarrow{\mathop{\longrightarrow}}
\def\opsim{\mathop{\sim}}

\def\fig#1#2{\includegraphics[height=#1]{#2}}
\def\figx#1#2{\includegraphics[width=#1]{#2}}


\title{ Real-Space Renormalization for disordered systems 
at the level of Large Deviations  } 


\author{ C\'ecile Monthus }
 \affiliation{Institut de Physique Th\'{e}orique, 
Universit\'e Paris Saclay, CNRS, CEA,
91191 Gif-sur-Yvette, France}

\begin{abstract}
The real-space renormalization procedures on hierarchical lattices have been much studied for many disordered systems in the past at the level of their typical fluctuations. In the present paper, the goal is to analyze instead the renormalization flows for the tails of probability distributions in order to extract the scalings of their large deviations and the tails behaviors of the corresponding rate functions. We focus on the renormalization rule for the ground-state energy of the Directed Polymer model in a random medium, and study the various renormalization flows that can emerge for the tails as a function of the tails of the initial condition.

\end{abstract}

\maketitle


\section{ Introduction} 

The theory of large deviation has a long history in mathematics
 (see the books \cite{ellisbook,deuschel,dembo,denHollander,saintflour,timo}
and references therein), in particular in the area of disordered systems (see the the books 
\cite{talagrand,bovier,comets}, the review \cite{zeitouni}  and references therein).
In physics, the explicit use of the large deviations framework is more recent but 
is nowadays recognized as the unifying language
 for equilibrium, non-equilibrium and dynamical systems
 (see the reviews \cite{oono,ellis,review_Touchette} and references therein).
In particular, this point of view has turned out to be essential 
to formulate the statistical physics approach of non-equilibrium dynamics
(see the reviews  \cite{derrida-lecture,harris_Schu,searles,harris,mft,lazarescu_companion,lazarescu_generic}
and the PhD Theses \cite{fortelle_thesis,vivien_Thesis,chetrite_Thesis,wynants_Thesis} 
 and the HDR Thesis \cite{chetrite_HDR}).

It is thus natural to revisit also classical and quantum disordered systems from the perspective of large deviations
\cite{c_largedevdisorder}. In particular, in the field of 
 real-space renormalization procedures for classical statistical physics models,
the focus of previous studies has been mostly the region of typical fluctuations around typical values,
but it is interesting to study now how their large deviations properties emerge from the renormalization flows.
In the present paper, we have chosen to focus on the renormalization rule for the intensive energy of the ground state of the Directed Polymer
on a hierarchical lattice depending on two integer parameters $A$ and $B$ (see section \ref{sec_diamond} for more details) : the new random variable $x_{n+1}$ at generation $(n+1)$ 
is obtained from $(AB)$ independent random variables $x_n^{(a,b)} $ of generation $n$ 
with $a=1,..,A$ and $b=1,..,B$
by the following maximum and sum operations
\begin{eqnarray}
x_{n+1} = \max \limits_{1 \leq b \leq B}
 \left( \frac{1}{A} \sum_{a=1}^A x_n^{(a,b)}  \right)
\label{recdiamond}
\end{eqnarray}
Our goal will be to study the renormalization flows for the tails $ x \to \pm \infty$ of the corresponding probability distribution ${\cal P}_n (x)  $ at generation $n$
as a function of the exponents $\alpha^{\pm} $ characterizing the exponential decays of the initial condition at generation $n=0$
\begin{eqnarray}
{\cal P}_0 (x) \oppropto_{x \to \pm \infty}  e^{- \lambda_0^{\pm} \vert x \vert^{\alpha^{\pm}} } 
\label{x0intro}
\end{eqnarray}

Besides its physical interpretation for the Directed Polymer model,
the RG rule of Eq. \ref{recdiamond} is also interesting on its own from the general point of view of probabilities,
because it mixes the basic operations 'sum over A independent variables' 
and 'maximum over B independent variables'.
Of course, the two following degenerate cases are very well-known:

(i) in the special case $B=1$, the variable 
\begin{eqnarray}
x_{n} =  \frac{1}{A} \sum_{a_1=1}^A x_{n-1}^{(a_1)}  
=  \frac{1}{A^2} \sum_{a_1=1}^A \sum_{a_2=1}^A x_{n-2}^{(a_1;a_2)}  =...
= \frac{1}{A^n} \sum_{a_1=1}^A ... \sum_{a_n=1}^A x_{0}^{(a_1;a_2;..;a_n)} 
\label{recdiamondb1}
\end{eqnarray}
reduces to the empirical average of $A^n$
independent variables $x_{0}^{(a_1;a_2;..;a_n)}  $ of generation $n=0$,
 which is the most studied problem in the whole history of probability.
The typical fluctuations are classified in terms of
the Gaussian distribution of the Central Limit Theorem
(see \cite{jona1,jona2,calvosum} for the renormalization point of view)
 and in terms of the L\'evy stable laws (when the variance does not exist).
While the standard theory for the large deviations of the empirical average
focuses on the case of symmetric large deviations \cite{oono,review_Touchette},
the case of asymmetric large deviations 
(with different scalings for rare values bigger or smaller than the typical value) have also attracted a lot of attention recently \cite{nagaev,evans2008,nina,godreche,evans2014,c_largedevasym}.
As recalled in Appendix \ref{app_convol}, the tails properties of the 
empirical average of Eq. \ref{recdiamondb1} strongly depend 
on the tail exponents $\alpha^{\pm} $ of the initial condition of Eq. \ref{x0intro}
with completely 
different regimes associated to compressed exponentials $\alpha^{\pm} >1$,
stretched exponentials $0<\alpha^{\pm} <1$ and simple exponentials $\alpha^{\pm} =1$.
For the more general problem of Eq. \ref{recdiamond}, one thus expects
that the tail exponents $\alpha^{\pm}  $ of the initial condition of Eq. \ref{x0intro}
will continue to play an essential role.

(ii) in the special case $A=1$, the variable
\begin{eqnarray}
x_{n} = \max \limits_{1 \leq b_1 \leq B} \left( x_{n-1}^{(b_1)}  \right)
=  \max \limits_{1 \leq b_1 \leq B ; 1 \leq b_2 \leq B} \left( x_{n-2}^{(b_1;b_2)}  \right)
= ...
= \max \limits_{1 \leq b_1 \leq B ;... ; 1 \leq b_n \leq B} \left( x_{0}^{(b_1;...;b_n)}  \right)
\label{recdiamonda1}
\end{eqnarray}
reduces to the empirical maximum of $B^n$
independent variables $x_{0}^{(b_1;...;b_n)}$ of generation $n=0$,
which is the basic problem in the field of Extreme Value Statistics \cite{Gum,Gal}.
 The typical fluctuations are classified in terms of
the three universality classes Gumbel-Fr\'echet-Weibull \cite{Gum,Gal}, with many applications in various physics domains (see the reviews \cite{mezard,clusel,fortin} and references therein)
and have been much studied from the renormalization perspective \cite{extreme1,extreme2,extreme3,extreme4,extreme5}.
The large deviations properties of the empirical maximum have been found to be asymmetric \cite{maxmath,vivo,c_largedevasym}, as a consequence of the following obvious asymmetry : 
an 'anomalously good' maximum requires only one anomalously good variable,
while an 'anomalously bad' maximum requires that all variables are anomalously bad.
This simple argument allows to understand why the large deviations will be also completely
different for the two tails $x \to \pm \infty$ in the more general problem of Eq. \ref{recdiamond}.

Since the symmetric and asymmetric large deviations properties
of these two special cases (i) and (ii) have been revisited in great detail recently in the companion paper \cite{c_largedevasym}, we will focus here on the non-degenerate cases $(A>1,B>1)$
where there is really an interplay between the maximum and the sum operations.
The paper is organized as follows.
In section \ref{sec_diamond}, we recall the origin of the RG rule of Eq. \ref{recdiamond}
for the Directed Polymer model on the hierarchical lattice of parameters $(A,B)$,
and we introduce the useful notations to analyze the renormalizations of the corresponding
probability distributions.
The various renormalization flows that can emerge for the two tails $x \to \pm\infty $
as a function of the exponents $\alpha^{\pm}$ of the tails of the initial condition (Eq \ref{x0intro})
are then discussed in the following sections with their consequences for the large deviations properties.
Section \ref{sec_aplusbig} describes the generic large deviation form with respect to the length $L_n=A^n$ that emerges for the positive tail $x \to +\infty$ 
 when the initial condition corresponds to some compressed exponential decay $\alpha^+>1$.
Similarly, section \ref{sec_amoinsbig} describes
the generic large deviation form with respect to the volume $L_n^d=A^{dn}$ 
that emerges for the negative tail $x \to -\infty$
when the initial condition
 corresponds to some compressed exponential decay $\alpha^->1$ . 
The anomalous large deviations properties that emerge when the initial condition decays only as a stretched exponential 
$0< \alpha^{\pm} <1$ are then discussed for the tails $x \to +\infty$ and $x \to -\infty$
 in sections  \ref{sec_aplussmall} and \ref{sec_amoinssmall} respectively.
Finally, the intermediate simple exponential decays $\alpha^+=1$ and $\alpha^-=1$ are considered in sections 
\ref{sec_aplus1} and \ref{sec_amoins1} respectively.
Our conclusions are summarized in Section \ref{sec_conclusion}.
The Appendix \ref{app_convol} contains a reminder on the tails properties 
of the empirical average of independent variables.

\section{ Real Space Renormalization at the level of large deviations}

\label{sec_diamond}

\subsection{ Hierarchical diamond lattice with two parameters $(A,B)$ }

\begin{figure}
\begin{center}
\includegraphics[width=16cm]{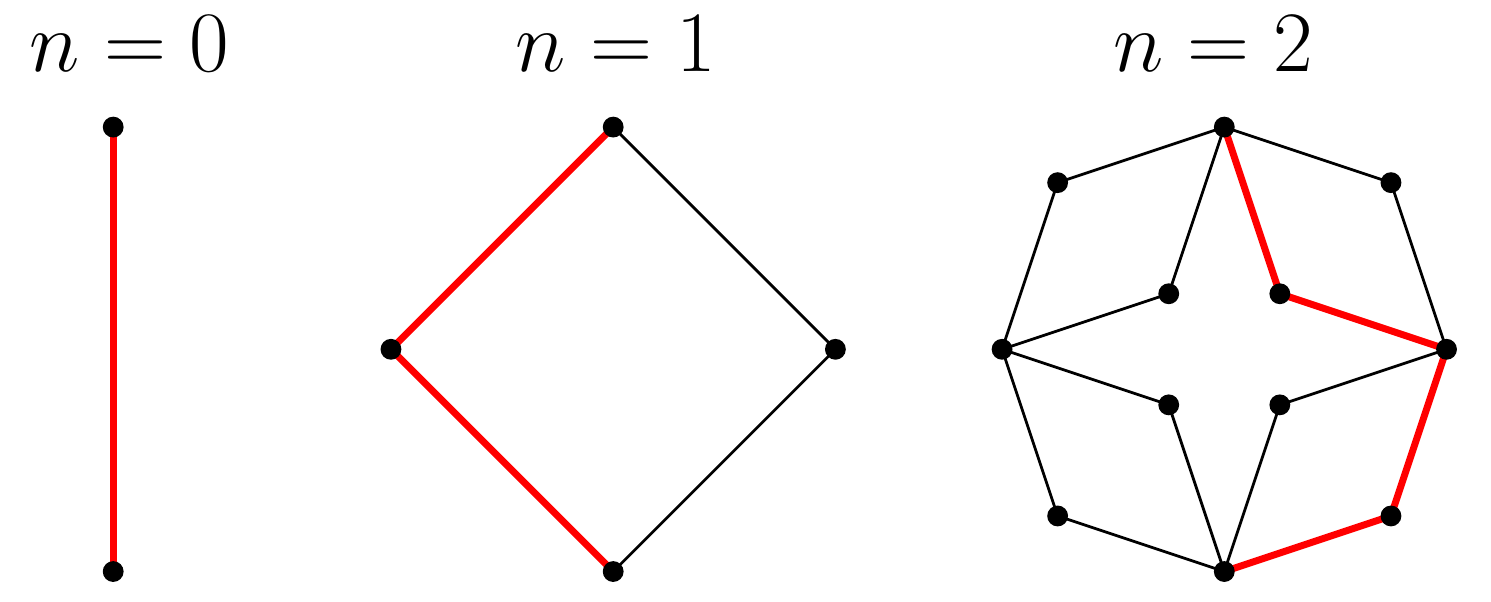}
\end{center}
\caption{ First generations $n=0,1,2$ of the hierarchical lattice of parameters $A=B=2$ :
the lengths $L_n=A^n=2^n$ between the two extreme points 
are given by $L_0=1$, $L_1=2$, $L_2=4$,
while the volumes $V_n=(AB)^n=4^n$ (total numbers of bonds) are given by
 $V_0=1$, $V_1=4$, $L_2=16$. An example of Directed Polymer of length $L_n$ between the two extreme points is shown in red for each generation.}
\label{figure}
\end{figure}

Among real-space renormalization procedures for classical statistical physics models
(see the reviews \cite{realspaceRG1,realspaceRG2,realspaceRG3} and references therein), 
Migdal-Kadanoff block renormalizations \cite{MKRG1,MKRG2} play a special role
because they can be considered in two ways, 
 either as approximate renormalization procedures on hypercubic lattices,
or as exact renormalization procedures on certain hierarchical lattices
\cite{berker,hierarchical1,hierarchical2}.
One of the most studied hierarchical lattice is the
diamond lattice which is constructed recursively
from the generation $n=0$
that contains a single bond of unit length $L_{n=0}=1$ 
by the following rule :
 the generation $n+1$ is made of $B$ branches, where each of these $B$ branches
 contains $A$ bonds of generation $n$ in series.
At generation $n$, the length $L_n$ between the two extreme sites is thus
\begin{eqnarray}
L_{n} && = A L_{n-1} = A^2 L_{n-2} = .. = A^n L_0= A^n
\label{ln}
\end{eqnarray}
while the volume $V_n$ (defined as the total number of bonds at generation $n$) 
grows as
\begin{eqnarray}
V_{n} && = (AB) V_{n-1} =(AB)^2 V_{n-2} ... = (AB)^n V_0= (AB)^n
\label{rec}
\end{eqnarray}
The effective fractal dimension $d$ that can be defined from the volume-length scaling
$V_n=L_n^d$ 
\begin{eqnarray}
d =\frac{ \ln(V_n) }{\ln (L_n) } = \frac{ \ln(AB) }{\ln A} = 1+ \frac{ \ln B }{\ln A}
\label{dliens}
\end{eqnarray}
allows to analyze the role of the dimensionality.
The special cases mentioned in the Introduction correspond to two extreme cases 
for the effective dimension :
the case (i) where $B=1$ corresponds to the dimension $d=1$,
and the lattice at generation $n$ reduces a series of $L_n$ bonds;
the case (ii) where $A=1$ corresponds to the dimension $d=\infty$,
because the length cannot grow and remains fixed to unity $L_n=A^n=1$,
while the volume $V_n=B^n$ grows, and the lattice at generation $n$ reduces 
to $B^n$ bonds in parallel. Apart from these two degenerate cases, 
the next simplest case $A=2=B$ corresponds to the effective dimension $d=2$,
and the first generations $n=0,1,2$ are shown on Figure \ref{figure} as example.

On these diamond lattices, many disordered models have been studied,
including the diluted Ising model \cite{diluted}, 
the ferromagnetic random Potts model \cite{Kin_Dom,Der_Potts,andelman,us_interface},
spin-glasses \cite{young1,young2,mckay,hartford,Gardnersg,bray1,bray2,moore,muriel1,muriel2,thill}
and the directed polymer model in a random medium
 \cite{Der_Gri,Coo_Der,Tim,roux,kardar1,kardar2,cao, tang,Muk_Bha,Bou_Sil,us_quadratic,us_tails}.
In this paper, we will focus only on the ground-state energy of this directed polymer model.

\subsection{ RG rules for the intensive ground state energy $x_n$ of the Directed Polymer Model}

At generation $n$, the number ${\cal N}_n$ of directed paths of length $L_n$
between the two extreme sites satisfies the recurrence
\begin{eqnarray}
{\cal N}_{n+1} =B ({\cal N}_{n}  )^A
\label{recchemin}
\end{eqnarray}
Taking into account the initial condition ${\cal N}_{n=0} =1$ at generation $n=0$,
one obtains the solution
\begin{eqnarray}
\ln {\cal N}_{n} && =\ln B + A \ln {\cal N}_{n-1}
= \ln B + A \left( \ln B + A \ln {\cal N}_{n-2} \right) = ...
\nonumber \\
&& = ( \ln B) \sum_{k=0}^{n-1} A^k = ( \ln B) \frac{A^n-1}{A-1}
= ( \ln B) \frac{L_n-1}{A-1}
\label{soluchemin}
\end{eqnarray}
that corresponds to the following configurational entropy per unit length
\begin{eqnarray}
s \equiv \lim_{n \to + \infty} \frac{ \ln {\cal N}_{n} }{L_n}  =  \frac{\ln B}{A-1}
\label{entropy}
\end{eqnarray}
which is finite for the non-degenerate cases $(A>1,B>1)$.

In the Model of the Directed Polymer in a random medium, a random energy $E_0$ 
is drawn independently for each bond of generation $n=0$.
At generation $n$, each directed path of length $L_n$
between the two extreme sites will collect
$L_n$ random energies of generation $n=0$,
 and the ground-state will correspond to the Directed path of minimum energy.
The hierarchical structure of the lattice yields that
the extensive ground state energy follows the closed renormalization rule \cite{Der_Gri}
\begin{eqnarray}
E_{n+1} = \min \limits_{1 \leq b \leq B}
 \left( \sum_{a=1}^A E_n^{(a,b)}  \right)
\label{rge}
\end{eqnarray}
where $ E_n^{(a,b)}$ are $(AB)$ independent energies of generation $n$.
In order to analyze the large deviations properties, it is more convenient to focus on the intensive 
variables that represent the ground-state energies per unit length (with a minus sign)
\begin{eqnarray}
x_n^{(a,b)} \equiv - \frac{E_n^{(a,b)}}{L_n} =  - \frac{E_n^{(a,b)}}{A^n} 
\label{xnrescal}
\end{eqnarray}
The RG rule of Eq. \ref{rge} then translates into the RG rule
\begin{eqnarray}
x_{n+1} \equiv - \frac{E_{n+1}}{A^{n+1}}  =\max \limits_{1 \leq b \leq B}
 \left(\frac{1}{A} \sum_{a=1}^A \left[- \frac{ E_n^{(a,b)} }{A^n} \right] \right)
=
 \max \limits_{1 \leq b \leq B}
 \left( \frac{1}{A} \sum_{a=1}^A x_n^{(a,b)}  \right)
\label{recdiamondbis}
\end{eqnarray}
already mentioned in Eq. \ref{recdiamond} of the Introduction.

\subsection{ RG rules for the probability distribution ${\cal P}_n(x) $ }

The RG rule of Eq. \ref{recdiamond} concerning random variables can be translated as follows for their probability distributions. If ${\cal P}_n(x)$ denotes the probability distribution of the independent intensive variables
 $x_n^{(a,b)} $ at generation $n$, the probability distribution ${\cal P}_{n+1}(x) $ at the next generation $(n+1)$
is then obtained via the two following steps \cite{Der_Gri} :

(1) the probability distribution ${\cal A}_n(x) $
of the $B$ independent empirical averages
\begin{eqnarray}
x_n^{(b)} \equiv \frac{1}{A} \sum_{a=1}^A x_n^{(a,b)} 
\label{xnempi}
\end{eqnarray}
is given by the convolution
\begin{eqnarray}
 {\cal A}_n(x)= \int_{-\infty}^{+\infty} dx_1 ... \int_{-\infty}^{+\infty} dx_A 
{\cal P}_n(x_1) ... {\cal P}_n(x_A) \delta\left( x- \frac{1}{A}\sum_{a=1}^A x_a   \right)
\label{Anconvolution}
\end{eqnarray}
The tails properties of this convolution $ {\cal A}_n(x)$ depend on the tails properties
of ${\cal P}_n(x) $ : as recalled in Appendix \ref{app_convol},
different regimes appear for compressed exponentials $\alpha^{\pm} >1$,
stretched exponentials $0<\alpha^{\pm} <1$ and simple exponentials $\alpha^{\pm} =1$.

(2) the probability distribution ${\cal P}_{n+1}(x) $ corresponds to the distribution of the maximum of $B$
independent variables $x_n^{(b)} $ of Eq. \ref{xnempi}
\begin{eqnarray}
 {\cal P}_{n+1}(x) 
= B  {\cal A}_n(x)\left[ \int_{-\infty}^x dx' {\cal A}_n(x') \right]^{B-1}
= B  {\cal A}_n(x)\left[ 1-\int_x^{+\infty} dx' {\cal A}_n(x') \right]^{B-1}
\label{dericumulXfull}
\end{eqnarray}
The two tails of ${\cal P}_{n+1}(x)  $ for $x \to \pm \infty$ are thus related to the tails of ${\cal A}_n(x) $ as follows
\begin{eqnarray}
 {\cal P}_{n+1}(x) && \opsimeq_{x \to +\infty} B  {\cal A}_n(x)
\nonumber \\
{\cal P}_{n+1}(x) && \opsimeq_{x \to -\infty} B  {\cal A}_n(x)\left[ \int_{-\infty}^x dx' {\cal A}_n(x') \right]^{B-1}
\label{dericumulX}
\end{eqnarray}
where one sees why the two tails $x \to \pm \infty$ will be governed by completely different mechanisms.

The goal of the present paper is to analyze the renormalization rules for the 
functions $f^{\pm}_n(x) $ that characterize the two tails $x \to \pm\infty $
\begin{eqnarray}
{\cal P}_{n}(x)  \opsimeq_{ x \to \pm\infty } e^{- f^{\pm}_n(x) }
\label{deffn}
\end{eqnarray}
in order to extract the large deviation properties for large $n$.

\subsection{ Link with the large deviations of the intensive ground-state energy }

The general expectation for the Directed Polymer model in a random medium of dimension $d$
is that the region of values bigger than the typical value ($x > x_{typ}$)
 should display a large deviation form with respect to the length $L_n$
\begin{eqnarray}
{\cal P}_n( x  )  && \oppropto_{L_n \to +\infty} e^{ -L_n I^+ (x)}  \ \ {\rm for } \ \ x > x^{typ}
\label{largedevp}
\end{eqnarray}
because an 'anomalously good' ground state energy requires only $L_n$ anomalously good bond energies along the polymer.
The region of values smaller than the typical value ($x < x_{typ}$)
 should display instead a large deviation form with respect to the volume $V_n=L_n^d$
\begin{eqnarray}
{\cal P}_n( x  )  && \oppropto_{L_n \to +\infty} e^{ -L_n^d I^- (x)}  \ \ {\rm for } \ \ x < x^{typ}
\label{largedevm}
\end{eqnarray}
because an 'anomalously bad' ground state energy requires $L_n^d$ bad bond energies in the sample.
This asymmetric large deviation form has been computed exactly for the Directed Polymer in dimension $d=1+1$
\cite{derrida_leb,dean_maj} that belongs to the Kardar-Parisi-Zhang universality class (see the various models and interpretations in the review \cite{gregory}).
Here our goal will be thus to derive this asymmetric large deviation form
for the hierarchical lattices of arbitrary parameters $(A,B)$,
and to compute explicitly the tails $x \to \pm\infty $ of the corresponding rate functions $I^{\pm} (x) $.

\subsection{ Special families of explicit solutions for the renormalization flows of the tails
$x \to \pm \infty$ }

While we will write the functional renormalization rules for the tail functions $f^{\pm}_n(x)$
of Eq. \ref{deffn}, the analysis of their general solutions in the infinite-dimensional space of all admissible tail functions
clearly goes beyond the goals of the present paper. We will instead focus on special families of explicit solutions
that appear for initial conditions of the following form
\begin{eqnarray}
{\cal P}_0 (x) \opsimeq_{x \to \pm \infty} 
K_0^{\pm} \vert x \vert^{\nu_0^{\pm}-1} e^{- \lambda_0^{\pm} \vert x \vert^{\alpha^{\pm}} } 
\label{x0}
\end{eqnarray}
As already stressed many times, the exponents $\alpha^{\pm}>0$ characterizing the leading exponential decays will play an essential role.
These parameters $\alpha^{\pm} $ will turned out to be conserved by the renormalization 
flow, while the other parameters 
may be renormalized, i.e. the tails at generation $n$ will be of the form
\begin{eqnarray}
{\cal P}_{n}(x)  \opsimeq_{ x \to \pm\infty } 
K_n^{\pm} \vert x \vert^{\nu_n^{\pm}-1} e^{- \lambda_n^{\pm} \vert x \vert^{\alpha^{\pm}} } 
\label{tailxn}
\end{eqnarray}
and will thus correspond to the following special form of the tail functions $f^{\pm}_n(x) $
\begin{eqnarray}
f^{\pm}_n(x)  \opsimeq_{ x \to \pm\infty } \lambda_n^{\pm} \vert x \vert^{\alpha^{\pm}} 
+ (1-\nu_n^{\pm} ) \ln \vert x \vert -\ln \left(K_n^{\pm}\right)  
\label{fspecial}
\end{eqnarray}
For each case labelled by the possible tail exponents $\alpha^{\pm} $, 
we will thus compute the solutions of 
the closed RG flows for the three other parameters $(\lambda_n^{\pm},\nu_n^{\pm},K_n^{\pm})$,
in order to extract the large deviations properties and the corresponding rate functions.

The following sections are devoted to the various renormalization flows 
that can emerge for the two tails $x \to \pm\infty $
as a function of the exponents $\alpha^{\pm} $ of the initial condition of Eq. \ref{x0} :
we will first consider the compressed exponential cases $\alpha^{\pm}>1$ that indeed lead to the expected large deviations of Eqs
\ref{largedevp} and \ref{largedevm}; we will then turn to the stretched exponential
cases $0< \alpha^{\pm}<1$ that lead to anomalous large deviations with respect to Eqs \ref{largedevp} and \ref{largedevm}; finally, we will discuss the intermediate cases $\alpha^{\pm}=1 $
that require a special analysis.


\section{ RG flow of the tail $x \to +\infty$ for the compressed exponential
cases $\alpha^+>1$ }

\label{sec_aplusbig}

\subsection{ Functional renormalization for the tail function $f_n^+(x) $ }

As recalled in Appendix \ref{app_convol}, 
when the tail function $f_n^+(x)$ of Eq. \ref{deffn}
satisfies the conditions of Eq. \ref{convex},
the tail of the distribution of the convolution $ {\cal A}_n(x)  $ of Eq. \ref{Anconvolution}
has been studied in detail in Ref \cite{frisch} and 
the output is the 'democratic' formula of Eq. \ref{frischa} 
\begin{eqnarray}
 {\cal A}_n(x)  \opsimeq_{x \to + \infty}  e^{ -A f^{+}_n(x) } \sqrt{A} \left( \frac{2 \pi} { (f^{+}_n)''(x) }  \right)^{\frac{A-1}{2} }
\label{frischan}
\end{eqnarray}

The tail $x \to +\infty$ of Eq. \ref{dericumulX} is simply given by
\begin{eqnarray}
{\cal P}_{n+1}(x) 
&& \opsimeq_{ x \to +\infty }  B {\cal A}_n(x)  
\opsimeq_{ x \to +\infty } B e^{ -A f^+_n(x) }  \sqrt{A} \left( \frac{2 \pi} { (f_n^+)''(x) }  \right)^{\frac{A-1}{2} }
\label{Xitertail}
\end{eqnarray}

The identification with ${\cal P}_{n+1}(x\to +\infty)  \simeq e^{- f_{n+1}^+(x)} $ of Eq. \ref{deffn}
 yields the functional RG rule for the tail function $f_n^+(x) $
\begin{eqnarray}
f^+_{n+1}(x) =  A  f^+_n(x) + (A-1) \ln \left( \sqrt{ \frac { (f_n^+)''(x) } {2 \pi} } \right) 
 -\ln \left( B  \sqrt{ A   }  \right)
\label{recfp}
\end{eqnarray}

\subsection{ Explicit solution of the RG flow for the special form of Eq. \ref{fspecial} when $\alpha^+>1$ }

The special form of Eq. \ref{fspecial} 
\begin{eqnarray}
f^+_n(x)  && \opsimeq_{ x \to +\infty } \lambda_n^+  x^{\alpha^+}+ (1-\nu_n^+ ) \ln x -\ln \left(K_n^+\right)  
\nonumber \\
(f^+_n)''(x)  && \opsimeq_{ x \to +\infty } \lambda_n^+ \alpha^+ (\alpha^+-1) x^{\alpha^+-2}+\frac{ (\nu_n^+ -1 ) }{x^2}
\opsimeq_{ x \to +\infty } \lambda_n^+ \alpha^+ (\alpha^+-1) x^{\alpha^+-2}
\label{fspecialp}
\end{eqnarray}
satisfies the conditions of Eq. \ref{convex} in the region $\alpha^+>1$, 
and 
remains closed under the functional RG flow of Eq. \ref{recfp}
with the following RG rules for the parameters
\begin{eqnarray}
\lambda_{n+1}^+ && = A \lambda_n^+
\nonumber \\
 \nu_{n+1}^+  && = A \left(  \nu_n^+-\frac{\alpha^+}{2} \right) + \frac{\alpha^+}{2}
\nonumber \\
\ln (K_{n+1}^+ ) && = A \ln (K_{n}^+ ) +(A-1) \ln \left( \sqrt{ \frac{2 \pi  }{\lambda_n^+\alpha^+(\alpha^+-1)} }\right)
+ \ln (B \sqrt{A})
\label{recspecial}
\end{eqnarray}

In terms of the initial condition at generation $n=0$,
the solution reads
\begin{eqnarray}
\lambda_n^+  && = A^n \lambda^+_0 
\nonumber \\
 \nu_n^+ && = A^n  \left(\nu_0^+-\frac{\alpha^+}{2} \right) + \frac{\alpha^+}{2} 
\nonumber \\
 \ln(K_n^+)
 && = A^n \left[
  \frac{ \ln B}{A-1} + \ln \left( K_0^+\sqrt{ \frac{2 \pi  }{\lambda^+_0 \alpha^+(\alpha^+-1)} }\right)
  \right]
 + \frac{n}{2} \ln A 
- \frac{ \ln B}{A-1} - \ln \left( \sqrt{ \frac{2 \pi  }{\lambda^+_0 \alpha^+(\alpha^+-1)} }\right)
\label{rgn}
\end{eqnarray}
Putting everything together, it is convenient to gather all the terms involving the length $L_n=A^n$
to obtain the final result for the tail function of Eq. \ref{fspecialp}
\begin{eqnarray}
f_n^+ (x) && \opsimeq_{x \to +\infty} A^n \left[\lambda_0^+ x^{\alpha^+ } 
 - \frac{ \ln B}{A-1} - \ln \left( K_0^+\sqrt{ \frac{2 \pi  }{\lambda^+_0\alpha^+(\alpha^+-1)} }\right)
 + \left(\frac{\alpha^+}{2}-\nu_0^+ \right) \ln x   \right]
 \nonumber \\
&& - \frac{n}{2} \ln A 
+ \frac{ \ln B}{A-1} +\ln \left( \sqrt{ \frac{2 \pi  }{\lambda^+_0\alpha^+(\alpha^+-1)} }\right)
+ \left(1-\frac{\alpha^+}{2} \right) \ln x 
\label{fpsolup}
\end{eqnarray}

\subsection{ Conclusion for the large deviations in the tail $x \to +\infty$ when $\alpha^+>1$   }

The RG solution of Eq. \ref{fpsolup} thus corresponds to the expected large deviation form 
with respect to the length $L_n=A^n$ of Eq. \ref{largedevp}.
In addition, the corresponding rate function $I^+(x) $ of Eq. \ref{largedevp}
displays the tail behavior
\begin{eqnarray}
I^+ (x) && \opsimeq_{x \to +\infty} 
\lambda_0^+ x^{\alpha^+ } 
 - \frac{ \ln B}{A-1} - \ln \left( K_0^+\sqrt{ \frac{2 \pi  }{\lambda^+_0\alpha^+(\alpha^+-1)} }\right)
 + \left(\frac{\alpha^+}{2}-\nu_0^+ \right) \ln x 
\label{tailratep}
\end{eqnarray}

\subsection{ Example with the Gaussian initial condition }

The special solution of Eq. \ref{fpsolup} will not contain the logarithmic terms in $(\ln x ) $
 for the initial conditions satisfying
\begin{eqnarray}
\alpha^+ && =2
 \nonumber \\
\nu_0^+&& =1
\label{gausspara}
\end{eqnarray}
It is thus interesting to consider the normalized Gaussian initial distribution at generation $n=0$
\begin{eqnarray}
{\cal P}_0(x) && =  K_0^+ e^{- \lambda_0^+x^2} 
\nonumber \\
K_0^+ && =\sqrt{ \frac{\lambda_0^+}{ \pi} }
\label{gauss0}
\end{eqnarray}
The special solution of Eq. \ref{fpsolup} then simplifies into
\begin{eqnarray}
f_n^+ (x) && \opsimeq_{x \to +\infty} A^n \left[\lambda_0^+ x^2
 - \frac{ \ln B}{A-1}   \right]
 - \frac{n}{2} \ln A 
+ \frac{ \ln B}{A-1} -\ln \left( \sqrt{ \frac{\lambda_0^+}{ \pi} } \right)
\label{fpsolupgauss}
\end{eqnarray}
and corresponds for the probability distribution to the tail
\begin{eqnarray}
{\cal P}_n(x) \opsimeq_{x \to +\infty} e^{-f_n^+ (x)} 
 \opsimeq_{x \to +\infty} \sqrt{ \frac{\lambda_0^+A^n}{\pi}  }
e^{ - A^n \left[\lambda_0^+ x^2 - \frac{ \ln B}{A-1}   \right] 
- \frac{ \ln B}{A-1}  }
= \sqrt{ \frac{\lambda_0^+ A^n}{\pi}  }
e^{ - A^n \left[ x^2 - (x_n^+)^2  \right] }
\label{pngauss}
\end{eqnarray}
with the parameter
\begin{eqnarray}
(x_{n}^+)^2  =  \frac{ \ln B}{A-1} \left( 1- \frac{1}{A^n} \right)
\label{xngauss}
\end{eqnarray}


\section{ RG flow of the tail $x \to -\infty$ for the compressed exponential cases $\alpha^->1$ }

\label{sec_amoinsbig}

\subsection{ Functional renormalization for the tail function $f_n^-(x) $ }

As recalled in Appendix \ref{app_convol}, 
when the tail function $f_n^-(x)$ of Eq. \ref{deffn}
satisfies the conditions of Eq. \ref{convex},
the tail of the distribution of the convolution $ {\cal A}_n(x)  $ of Eq. \ref{Anconvolution}
 is given by the 'democratic' formula of Eq. \ref{frischa} 
\begin{eqnarray}
 {\cal A}_n(x)  \opsimeq_{x \to - \infty}  e^{ -A f^{-}_n(x) } \sqrt{A} \left( \frac{2 \pi} { (f^{-}_n)''(x) }  \right)^{\frac{A-1}{2} }
\label{frischabis}
\end{eqnarray}
As a consequence, the corresponding cumulative distribution displays the tail
\begin{eqnarray}
\int_{-\infty}^x dx' {\cal A}_n(x') \opsimeq_{x \to -\infty} 
\int_{-\infty}^x dx'  e^{ -A f^{-}_n(x') } \sqrt{A} \left( \frac{2 \pi} { (f^{-}_n)''(x') }  \right)^{\frac{A-1}{2} }
\opsimeq_{x \to -\infty} 
 e^{ -A f^{-}_n(x) } \frac{ \sqrt{A} } { A [- (f^{-}_n)'(x) ] } \left( \frac{2 \pi} { (f^{-}_n)''(x) }  \right)^{\frac{A-1}{2} }
\label{cumulXm}
\end{eqnarray}
So the tail $x \to -\infty$ of Eq. \ref{dericumulX} is given by
\begin{eqnarray}
{\cal P}_{n+1}(x) 
&& \opsimeq_{ x \to -\infty }  B {\cal A}_n(x)  \left[ \int_{-\infty}^x dx' {\cal A}_n(x') \right]^{B-1}
\nonumber \\
&& 
\opsimeq_{x \to -\infty} B A [- (f^{-}_n)'(x) ]  
\left[  e^{ -A f^{-}_n(x) } \frac{ \sqrt{A} } { A [- (f^{-}_n)'(x) ] }
 \left( \frac{2 \pi} { (f^{-}_n)''(x) }  \right)^{\frac{A-1}{2} } \right]^B
\label{Xtailm}
\end{eqnarray}

The identification of the tail ${\cal P}_{n+1}(x \to -\infty)  \simeq e^{- f_{n+1}^-(x)} $ of Eq. \ref{deffn}
 yields the functional RG rule for the tail function $f_n^-(x) $
\begin{eqnarray}
f^-_{n+1}(x) =  AB   f^-_n(x)  +(A-1) B \ln \left( \sqrt{ \frac { (f^{-}_n)''(x) } {2 \pi } }\right) 
+ (B-1) \ln [ - (f^{-}_n)'(x) ]
+ \left( \frac{B}{2} -1\right) \ln A 
- \ln B
\label{recfm}
\end{eqnarray}

\subsection{ Explicit solution of the RG flow for the special form of Eq. \ref{fspecial} }

The special form of Eq. \ref{fspecial} 
\begin{eqnarray}
f^-_n(x)  && \opsimeq_{ x \to -\infty } \lambda_n^-  (-x)^{\alpha^-}+ (1-\nu_n^- ) \ln (-x) -\ln \left(K_n^-\right)  
\nonumber \\
(f^-_n)'(x)  && \opsimeq_{ x \to - \infty } - \lambda_n^- \alpha^- (-x)^{\alpha^--1}+ \frac{1-\nu_n^-}{x}
\nonumber \\
(f^-_n)''(x)  && \opsimeq_{ x \to - \infty } \lambda_n^- \alpha^- (\alpha^--1) (-x)^{\alpha^--2}+\frac{ (\nu_n^- -1 ) }{x^2}
\opsimeq_{ x \to - \infty } \lambda_n^- \alpha^- (\alpha^--1) (-x)^{\alpha^--2}
\label{fspecialm}
\end{eqnarray}
satisfies the conditions of Eq. \ref{convex} in the region $\alpha^->1$,
and remains closed under the functional RG flow of Eq. \ref{recfm}
with the following RG rules for the parameters
\begin{eqnarray}
\lambda_{n+1}^- && = AB \lambda_n^-
\label{recspecialm} \\
 \nu_{n+1}^-  && = AB   \nu_n^-   +\frac{\alpha^-}{2}(2-B-AB)
\nonumber \\
\ln (K_{n+1}^- ) && = AB \ln (K_{n}^- )- (AB-1) \ln(\lambda_n^-)
 +(A-1) B\ln \left( \sqrt{ \frac{2 \pi  }{\alpha^-(\alpha^--1)} }\right)
 - (B-1) \ln(\alpha^-) -  (B-2)\frac{\ln A}{2}
+ \ln (B)
\nonumber
\end{eqnarray}

In terms of the initial condition at generation $n=0$,
the solution reads
\begin{eqnarray}
\lambda_n^-  && = (AB)^n  \lambda_0^-
\nonumber \\
 \nu_n^- && = (AB)^n  \left( \nu_0^-  - \frac{\alpha^-}{2} \omega  \right) 
 + \frac{\alpha^-}{2} \omega
\nonumber \\
 \ln(K_n^-) && = (AB)^n \left[   \ln(K_0^-) +v  \right]
 + \frac{n}{2} \omega \ln (AB) 
- v  
\label{rgnm}
\end{eqnarray}
where we have introduced the notation
\begin{eqnarray}
\omega && \equiv 1+\frac{B-1}{AB-1} 
\nonumber \\
v && \equiv - \ln( \lambda_0^- ) +
\frac{(A-1)B  }{ AB-1} \left[ \frac{ \ln \sqrt{B} }{AB-1  }+  \ln \left( \sqrt{ \frac{2 \pi  }{\alpha^-(\alpha^--1)} }\right) \right]
-\frac{ B-1 }{ AB-1} \left[ \frac{AB}{AB-1}  \ln \sqrt{ A } + \ln(\alpha^- ) \right]
\label{defv}
\end{eqnarray}

Putting everything together, the tail function of Eq. \ref{fspecialm}
reads 
\begin{eqnarray}
f_n^- (x) \opsimeq_{x \to -\infty}  
&& (AB)^n \left[ \lambda_0^- \vert x \vert^{\alpha^- }  -\ln(K_0^-) -v 
+ \left( \frac{\alpha^-}{2} \omega -\nu_0^- \right)  \ln \vert x \vert  \right]
\nonumber \\ 
&&
 - \frac{n}{2} \omega \ln (AB) 
+ v  + \left( 1-\frac{\alpha^-}{2} \omega\right)  \ln \vert x \vert 
\label{solufnm}
\end{eqnarray}

\subsection{ Conclusion for the large deviations in the tail $x \to -\infty$ when $\alpha^->1$ }

The RG solution of Eq. \ref{solufnm} thus corresponds to the expected large deviation form 
with respect to the volume $V_n=L_n^d=(AB)^n$ of Eq. \ref{largedevm}.
The corresponding rate function $I^-(x) $ of Eq. \ref{largedevm}
displays the tail behavior
\begin{eqnarray}
I^- (x) && \opsimeq_{x \to -\infty} 
\lambda_0^- \vert x \vert^{\alpha^- } 
+ \left( \frac{\alpha^-}{2} \omega -\nu_0^- \right)  \ln \vert x \vert 
 -\ln(K_0^-) -v 
\nonumber \\
&& = \lambda_0^- \vert x \vert^{\alpha^- } 
+ \left[ \frac{\alpha^-}{2} \left(1+\frac{B-1}{AB-1}   \right) -\nu_0^- \right]  \ln \vert x \vert 
\label{tailratem} \\
&& +\ln( \lambda_0^- ) 
 -\ln(K_0^-) 
-\frac{(A-1)B  }{ AB-1} \left[ \frac{ \ln \sqrt{B} }{AB-1  }+  \ln \left( \sqrt{ \frac{2 \pi  }{\alpha^-(\alpha^--1)} }\right) \right]
+\frac{ B-1 }{ AB-1} \left[ \frac{AB}{AB-1}  \ln \sqrt{ A } + \ln(\alpha^- ) \right]
\nonumber
\end{eqnarray}


\section{ RG flow of the tail $x \to +\infty$ for the stretched exponential cases $0<\alpha^+<1$ }

\label{sec_aplussmall}

\subsection{ Functional renormalization for the tail function $f_n^+(x) $ }

As recalled in Appendix \ref{app_convol}, 
the tail of the distribution of the convolution $ {\cal A}_n(x)  $ of Eq. \ref{Anconvolution}
is then given by the 'monocratic formula' of Eq. \ref{monocratic}
\begin{eqnarray}
 {\cal A}_n(x)  \opsimeq_{x \to + \infty}  A^2e^{ - f_n^+(Ax) } 
\label{monocraticp}
\end{eqnarray}
so the tail $x \to +\infty$ of Eq. \ref{dericumulX} becomes
\begin{eqnarray}
{\cal P}_{n+1}(x) 
&& \opsimeq_{ x \to +\infty }  B {\cal A}_n(x)  
\opsimeq_{ x \to +\infty } B A^2e^{ - f_n^+(Ax) } 
\label{Xitertails}
\end{eqnarray}

The identification with ${\cal P}_{n+1}(x \to +\infty)  \simeq e^{- f_{n+1}^+(x)} $ of Eq. \ref{deffn}
 yields the functional RG rule for the tail function $f_n^+(x) $
\begin{eqnarray}
f^+_{n+1}(x) =    f^+_n(A x)  -\ln \left( B A^2 \right)
\label{recfps}
\end{eqnarray}
instead of the functional RG rule of Eq. \ref{recfp}.

\subsection{ Explicit solution of the RG flow for the special form of Eq. \ref{fspecial} when $0<\alpha^+<1$}

The special form of Eq. \ref{fspecial} 
\begin{eqnarray}
f^+_n(x)  && \opsimeq_{ x \to +\infty } \lambda_n^+  x^{\alpha^+}+ (1-\nu_n^+ ) \ln x -\ln \left(K_n^+\right)  
\label{fspecialpbis}
\end{eqnarray}
remains closed for the functional RG rule of Eq. \ref{recfps}
 with the following RG rules for the parameters
\begin{eqnarray}
\lambda_{n+1}^+ && = A^{\alpha^+} \lambda_n^+
\nonumber \\
 \nu_{n+1}^+  && =   \nu_n^+
\nonumber \\
\ln (K_{n+1}^+ ) && =  \ln (K_{n}^+ ) +(\nu_n^+ +1) \ln A + \ln B
\label{recspecialbis}
\end{eqnarray}

In terms of the initial condition at generation $n=0$,
the solution reads
\begin{eqnarray}
\lambda_n^+  && = A^{n \alpha^+ }  \lambda^+_0 
\nonumber \\
 \nu_n^+ && = \nu_0^+
\nonumber \\
 \ln(K_n^+)  && = \ln ( K_0^+) + n \left[ (\nu^+_0+1) \ln A + \ln B \right]
\label{rgnbis}
\end{eqnarray}

Putting everything together, the tail function of Eq. \ref{fspecialpbis} reads
\begin{eqnarray}
f_n^+ (x) && \opsimeq_{x \to +\infty} A^{n \alpha^+ }  \lambda^+_0   x^{\alpha^+}+ (1-\nu_0^+ ) \ln x 
 -\ln ( K_0^+) - n \left[ (\nu^+_0+1) \ln A + \ln B \right]
\label{fpsolups}
\end{eqnarray}

\subsection{ Conclusion for the anomalous large deviations in the tail $x \to +\infty$ when $0<\alpha^+<1$ }

The solution of Eq. \ref{fpsolups} thus corresponds to the following anomalous large deviation form 
with respect to the length $L_n=A^n$ 
\begin{eqnarray}
{\cal P}_n( x  )  && \oppropto_{L_n \to +\infty} e^{ -L_n^{\alpha^+} J^+ (x)}  \ \ {\rm for } \ \ x \geq x^{typ}
\label{largedevpbis}
\end{eqnarray}
instead of the standard form of Eq. \ref{largedevp}.
The corresponding rate function $J^+(x) $ displays the tail behavior
\begin{eqnarray}
J^+ (x) && \opsimeq_{x \to +\infty}  \lambda_0^+ x^{\alpha^+ } 
\label{tailratepbis}
\end{eqnarray}


\section{ RG flow of the tail $x \to -\infty$ for the stretched exponential cases $0<\alpha^-<1$ }

\label{sec_amoinssmall}

\subsection{ Functional renormalization for the tail function $f_n^-(x) $ }

As recalled in Appendix \ref{app_convol}, 
the tail of the distribution of the convolution $ {\cal A}_n(x)  $ of Eq. \ref{Anconvolution}
is then given by the 'monocratic formula' of Eq. \ref{monocratic}
\begin{eqnarray}
 {\cal A}_n(x)  \opsimeq_{x \to - \infty}  A^2e^{ - f_n^-(Ax) } 
\label{monocraticm}
\end{eqnarray}
The corresponding cumulative distribution displays the tail
\begin{eqnarray}
\int_{-\infty}^x dx' {\cal A}_n(x') \opsimeq_{x \to -\infty} 
\int_{-\infty}^x dx'  A^2e^{ - f_n^-(Ax) } 
\opsimeq_{x \to -\infty}   \frac{ A } {  [- (f^{-}_n)'(Ax) ] } e^{ - f_n^-(Ax) } 
\label{cumulXm1}
\end{eqnarray}
and leads to the following result for the tail $x \to -\infty$ of Eq. \ref{dericumulX} 
\begin{eqnarray}
{\cal P}_{n+1}(x) 
&& \opsimeq_{ x \to -\infty }  B {\cal A}_n(x)  \left[ \int_{-\infty}^x dx' {\cal A}_n(x') \right]^{B-1}
 \opsimeq_{ x \to -\infty } \frac{B A^{B+1} }{  [- (f^{-}_n)'(Ax) ]^{B-1} }  e^{ - B f_n^-(Ax) } 
\label{Xtailm1}
\end{eqnarray}
The identification with ${\cal P}_{n+1}(x \to -\infty)  \simeq e^{- f_{n+1}^-(x)} $ of Eq. \ref{deffn}
 yields the functional RG rule for the tail function $f_n^-(x) $
\begin{eqnarray}
f^-_{n+1}(x) =  B  f^-_n(A x)  + (B-1) \ln [- (f^{-}_n)'(Ax) ]- (B+1) \ln A -\ln B
\label{recfms}
\end{eqnarray}
instead of the functional RG rule of Eq. \ref{recfm}.

\subsection{ Explicit solution of the RG flow for the special form of Eq. \ref{fspecial} for $0<\alpha^-<1$}

The special form of Eq. \ref{fspecial} 
\begin{eqnarray}
f^-_n(x)  && \opsimeq_{ x \to -\infty } \lambda_n^-  (-x)^{\alpha^-}+ (1-\nu_n^- ) \ln (-x) -\ln \left(K_n^-\right)  
\nonumber \\
(f^{-}_n)'(x) && \opsimeq_{ x \to -\infty } - \lambda_n^- \alpha^- (-x)^{\alpha^- -1}+ \frac{1-\nu_n^- }{x} \opsimeq_{ x \to -\infty } - \lambda_n^- \alpha^- (-x)^{\alpha^- -1}
\label{fspecialmbis}
\end{eqnarray}
remains closed for the functional RG rule of Eq. \ref{recfms}
with the following RG rules for the parameters
\begin{eqnarray}
\lambda_{n+1}^- && = B A^{\alpha^-} \lambda_n^-
\nonumber \\
 \nu_{n+1}^-  && = B   \nu_n^- - (B-1) \alpha^- 
\nonumber \\
\ln (K_{n+1}^- ) && = B \ln (K_{n}^- ) -(B-1) \ln(\alpha^-\lambda_n^- ) 
 +\left[ B(\nu_n^- +1) -(B-1) \alpha^-\right] \ln A + \ln B
\label{recspecialbism}
\end{eqnarray}

In terms of the initial condition at generation $n=0$,
the solution reads
\begin{eqnarray}
\lambda_n^-  && = \left( B A^{ \alpha^- }\right)^n  \lambda^-_0 
\label{rgnmbis} \\
 \nu_n^- && =  B^n (\nu_0^- -\alpha^- ) + \alpha^-
\nonumber \\
 \ln(K_n^-)  && = B^n \left[\ln(K_0^-) 
+ n (\nu_0^- -\alpha^- ) \ln A - \ln( \lambda^-_0 \alpha^-) 
+ \frac{B}{B-1} \ln A 
 \right]
+ n \ln(B A^{ \alpha^- } )
+ \ln( \lambda^-_0 \alpha^-) 
- \frac{B}{B-1}  \ln A 
\nonumber
\end{eqnarray}

Putting everything together, the tail function of Eq. \ref{fspecialmbis} reads
\begin{eqnarray}
f_n^- (x) && \opsimeq_{x \to - \infty}
\left( B A^{ \alpha^- }\right)^n  \lambda^-_0  \vert x \vert^{\alpha^-}
-B^n \left[  (\nu_0^- -\alpha^- )  \vert x \vert +     \ln(K_0^-) + n (\nu_0^- -\alpha^- ) \ln A - \ln( \lambda^-_0 \alpha^-) 
+ \frac{B}{B-1} \ln A 
 \right]
\nonumber \\ &&
- n \ln(B A^{ \alpha^- } )
 +(1- \alpha^- ) \vert x \vert
- \ln( \lambda^-_0 \alpha^-) 
+ \frac{B}{B-1}  \ln A 
\label{fpsolum}
\end{eqnarray}

\subsection{ Conclusion for the anomalous large deviations in the tail $x \to -\infty$ when $0<\alpha^-<1$ }

The solution of Eq. \ref{fpsolum}
 thus corresponds to the following anomalous large deviation form in $\left( B A^{ \alpha^- }\right)^n = L_n^{d-1+\alpha^-} $ 
\begin{eqnarray}
{\cal P}_n( x  )  && \oppropto_{L_n \to +\infty} e^{ -L_n^{d-1+\alpha^-} J^- (x)}  \ \ {\rm for } \ \ x \leq x^{typ}
\label{largedevpbism}
\end{eqnarray}
instead of the standard form of Eq. \ref{largedevm}.
The corresponding rate function $J^-(x) $ displays the tail behavior
\begin{eqnarray}
J^- (x) && \opsimeq_{x \to -\infty}  \lambda^-_0  \vert x \vert^{\alpha^-}
\label{tailratepbism}
\end{eqnarray}


\section{ RG flow of the tail $x \to +\infty$ for the intermediate cases $\alpha^+=1$  }

\label{sec_aplus1}

\subsection{ Explicit solution of the RG flow for the special form of Eq. \ref{fspecial} for $\alpha^+=1$ and $\nu_{0}^+>0 $}

In this section, we wish to analyze the closed RG flow for the special form of Eq. \ref{tailxn} when $\alpha^+=1$
\begin{eqnarray}
 {\cal P}_{n}(x) \opsimeq_{x \to + \infty} K_n^+ x ^{\nu^+_n-1} e^{- \lambda_n^+ x } 
\label{xnalpha1}
\end{eqnarray}
As explained in the Appendix \ref{app_convol}, the tail of the convolution of Eq. \ref{Anconvolution}
is then given by Eq. \ref{ana1} if $\nu_n^+>0 $
\begin{eqnarray}
 {\cal A}_n(x)\opsimeq_{x \to + \infty} \frac{ A^{A \nu_n^+ } 
\left[ K_n^+ \Gamma(\nu^+_n)\right]^A }{\Gamma(A \nu^+_n)} x^{A \nu_n^+-1}e^{- A \lambda_n^+ x }
\label{ana1n}
\end{eqnarray}
Then Eq. \ref{dericumulX} yields that the tail at generation $(n+1)$ reads
\begin{eqnarray}
{\cal P}_{n+1}(x) 
 \opsimeq_{ x \to +\infty }  B {\cal A}_n(x)  
 \opsimeq_{ x \to +\infty }  B 
\frac{ A^{A \nu_n^+ } 
\left[ K_n^+ \Gamma(\nu^+_n)\right]^A }{\Gamma(A \nu^+_n)} x^{A \nu_n^+-1}e^{- A \lambda_n^+ x }
\label{Xitertaila1}
\end{eqnarray}
The identification with the notations of Eq. \ref{xnalpha1} at generation $(n+1)$
leads to the following RG rules for the parameters
\begin{eqnarray}
\lambda_{n+1}^+ && = A \lambda_n^+
\nonumber \\
\nu_{n+1}^+ && = A \nu_n^+
\nonumber \\
\ln(K_{n+1}^+)  && = 
A \left[ \ln(K_n^+) + \ln \left( \Gamma(\nu^+_n) \right) + \nu_n^+ \ln A \right]
 - \ln \left( \Gamma(A \nu^+_n) \right)+ \ln(B)
\label{reca1}
\end{eqnarray}
Taking into account the initial condition at generation $n=0$,
the solution reads
\begin{eqnarray}
\lambda_{n}^+ && = A^n \lambda_0^+ 
\nonumber \\
\nu_{n}^+ && = A^n \nu_0^+ 
\nonumber \\
\ln( K_{n}^+) && = A^n \left[ \frac{\ln B}{A-1} +  n  \nu^+ \ln A +\ln(K_0^+) + \ln( \Gamma( \nu^+)) \right]
-\ln \left( \Gamma( A^n \nu^+) \right)
-\frac{1}{A-1} \ln B
\label{solureca1}
\end{eqnarray}
so this solution satisfies the validity condition $\nu_{n}^+>0 $ for any $n$ if the initial condition does $\nu_{0}^+>0 $.

Putting everything together, the tail function $f_n(x)$ of Eq. \ref{fspecial} reads
\begin{eqnarray}
f^{+}_n(x) && \opsimeq_{ x \to +\infty } \lambda_n^{+} x
+ (1-\nu_n^{+} ) \ln x -\ln \left(K_n^{+}\right)  
\label{fspeciala1} \\
&& \opsimeq_{ x \to +\infty }
A^n \left[\lambda_0^+  x -  \nu_0^+   \ln x  - \frac{\ln B}{A-1} -  n  \nu^+_0 \ln A - \ln(K_0^+) - \ln( \Gamma( \nu^+_0)) \right]
+ \ln x
+\ln \left( \Gamma( A^n \nu^+_0) \right)
+\frac{1}{A-1} \ln B
\nonumber
\end{eqnarray}

\subsection{ Conclusion for the large deviations in the tail $x \to +\infty$ for $\alpha^+=1$ and $\nu_{0}^+>0 $ }

To extract the large deviation form from the solution of Eq. \ref{fspeciala1},
one needs to use the Stirling formula for the Gamma function of $z=A^n \nu^+_0 \gg 1$
\begin{eqnarray}
\Gamma( A^n \nu^+_0) \opsimeq_{n \gg 1} 
\sqrt{ 2 \pi  } ( A^n \nu^+_0)^{ A^n \nu^+_0- \frac{1}{2}}
e^{-A^n \nu^+_0} 
\label{strirling}
\end{eqnarray}
Plugging its logarithm 
\begin{eqnarray}
\ln\left( \Gamma( A^n \nu^+_0) \right)\opsimeq_{n \gg 1} 
  A^n \left[ n \nu^+_0\ln A +\nu^+_0 \ln(\nu^+_0) - \nu^+_0\right]
+ \ln(\sqrt{ 2 \pi  })  - \frac{1}{2} \left( n \ln A + \ln(\nu^+_0) \right)
\label{lnstrirling}
\end{eqnarray}
into Eq. \ref{fspeciala1} yields to the standard large deviation form with respect to the length $L_n=A^n$ of Eq. \ref{largedevp}
and the corresponding rate function $I^+(x) $ displays the tail behavior
\begin{eqnarray}
I^+(x) \opsimeq_{ x \to +\infty }
\lambda_0^+  x   - \frac{\ln B}{A-1} 
 - \ln(K_0^+) - \ln( \Gamma( \nu^+_0)) 
 +\nu^+_0 \ln(\nu^+_0) - \nu^+_0  -  \nu_0^+   \ln x 
\label{xa1rate}
\end{eqnarray}
instead of Eq. \ref{tailratep}.


\section{ RG flow of the tail $x \to -\infty$ for the intermediate cases $\alpha^-=1$  }

\label{sec_amoins1}

\subsection{ Explicit solution of the RG flow for the special form of Eq. \ref{fspecial} for $\alpha^-=1$ and $\nu_{0}^- \geq \frac{B-1}{AB-1}  $}

In this section, we wish to analyze the closed RG flow for the special form of Eq. \ref{tailxn} when $\alpha^-=1$
\begin{eqnarray}
 {\cal P}_{n}(x) \opsimeq_{x \to - \infty} K_n^- \vert x \vert^{\nu^-_n-1} e^{- \lambda_n^- \vert x \vert }
\label{xnalpha1m}
\end{eqnarray}
As explained in the Appendix \ref{app_convol}, the tail of the convolution of Eq. \ref{Anconvolution}
is then given by the analog of Eq. \ref{ana1} if $\nu_n^->0 $
\begin{eqnarray}
 {\cal A}_n(x)\opsimeq_{x \to - \infty} \frac{ A^{A \nu_n^- } 
\left[ K_n^- \Gamma(\nu^-_n)\right]^A }{\Gamma(A \nu^-_n)} \vert x \vert^{A \nu_n^--1}e^{- A \lambda_n^- \vert x \vert }
\label{ana1m}
\end{eqnarray}
Then the corresponding cumulative distribution displays the tail
\begin{eqnarray}
 \int_{-\infty}^x dx' {\cal A}_n(x') \opsimeq_{x \to - \infty} \frac{ A^{A \nu_n^- } 
\left[ K_n^- \Gamma(\nu^-_n)\right]^A }{  A \lambda_n^-\Gamma(A \nu^-_n)} \vert x \vert^{A \nu_n^--1}e^{- A \lambda_n^- \vert x \vert }
\label{ana1mcumul}
\end{eqnarray}
As a consequence, the tail at generation $(n+1)$
of Eq. \ref{dericumulX} reads
\begin{eqnarray}
 {\cal P}_{n+1}(x) 
= B  {\cal A}_n(x)\left[ \int_{-\infty}^x dx' {\cal A}_n(x') \right]^{B-1}
\opsimeq_{x \to - \infty} \frac{B}{[ A \lambda_n^-]^{B-1}  } 
\left[  \frac{ A^{A \nu_n^- } 
\left[ K_n^- \Gamma(\nu^-_n)\right]^A }{\Gamma(A \nu^-_n)} \vert x \vert^{A \nu_n^--1}e^{- A \lambda_n^- \vert x \vert } \right]^{B}
\label{dericumulXa1m}
\end{eqnarray}
The identification with the notations of Eq. \ref{xnalpha1m} at generation $(n+1)$
leads to the following RG rules for the parameters
\begin{eqnarray}
\lambda_{n+1}^- && = AB \lambda_n^-
\nonumber \\
\nu_{n+1}^- && = AB \nu_n^- -(B-1)
\nonumber \\
\ln( K_{n+1}^-) && = 
AB  \ln(K_n^- )
+ B \left[ A \ln \left( \Gamma(\nu^-_n)\right)  -  \ln \left(\Gamma(A \nu^-_n) \right) \right]
+\nu_n^- AB\ln A   
- (B-1) \left[ \ln (\lambda_n^-) + \ln A \right]
 + \ln B 
\label{reca1m}
\end{eqnarray}

Taking into account the initial condition at generation $n=0$,
the solution reads
\begin{eqnarray}
\lambda_{n}^- && =   (AB)^n\lambda_0^- 
\nonumber \\
\nu_{n}^- && =  (AB)^n \left[ \nu^-_0 - \frac{B-1}{AB-1}   \right]  + \frac{B-1}{AB-1}  
\nonumber \\
\ln(K_{n}^-) && =  (AB)^n \left[ \ln(K_{0}^-) 
+n \left( \nu^-_0 - \frac{B-1}{AB-1} \right) \ln A
 + \frac{B(A-1)}{(AB-1)^2} \ln B
- \frac{B-1}{AB-1} \ln(\lambda_0^- )
 \right] 
\nonumber \\ &&
+n \frac{B-1}{(AB-1)}   \ln(AB)
-\frac{B(A-1)}{(AB-1)^2} \ln B
+ \frac{B-1}{AB-1} \ln(\lambda_0^- )
\nonumber \\ &&
+ \sum_{k=0}^{n-1} (AB)^{n-1-k} B \left[ A \ln \left( \Gamma(\nu^-_k)\right)  -  \ln \left(\Gamma(A \nu^-_k) \right) \right]
\label{solureca1m}
\end{eqnarray}
so this solution satisfies the validity condition $\nu_{n}^->0 $ for any $n$ if the initial condition satisfies 
$\nu_{0}^- \geq \frac{B-1}{AB-1} $.

\subsection{ Conclusion for the large deviations in the tail $x \to -\infty$ for $\alpha^-=1$ and $\nu_{0}^-> \frac{B-1}{AB-1}$ }

To extract the large deviation form from the solution of Eq. \ref{solureca1m},
one needs to use the Stirling formula for $\Gamma(\nu^-_k) $ and $\Gamma(A \nu^-_k) $
to obtain the asymptotic behavior of the difference
\begin{eqnarray}
 \left[ A \ln \left( \Gamma(\nu^-_k)\right)  -  \ln \left(\Gamma(A \nu^-_k) \right) \right]
\opsimeq_{k \gg 1}
&&  - (AB)^k A \left( \nu^-_0 - \frac{B-1}{AB-1} \right) \ln A
-k \frac{A-1}{2} \ln (AB)
\nonumber \\ &&+ \left[ (A-1) \ln (\sqrt{2 \pi} ) - \frac{A-1}{2}  \ln \left( \nu^-_0 - \frac{B-1}{AB-1} \right)
+ \left(  \frac{A-1}{AB-1}  - \frac{1}{2} \right) \ln A
\right]
\label{diffstirling}
\end{eqnarray}
As a consequence, the leading terms of order $(AB)^n$ in the solution $\ln(K_{n}^-) $ of Eq \ref{solureca1m} is given by
\begin{eqnarray}
\ln(K_{n}^-) 
 && 
\opsimeq_{n \gg 1}  (AB)^n \left[ \ln(K_{0}^-) 
   -  \frac{B}{2(AB-1)} \ln A
- \frac{B-1}{AB-1} \ln(\lambda_0^- )
+\frac{B(A-1)}{AB-1} \left( 
\frac{\ln (AB)}{2 (AB-1)} 
+ \ln \left(\sqrt{ \frac{ 2 \pi } {\nu^-_0 - \frac{B-1}{AB-1}  } } \right)
\right)
\right]
\nonumber \\
&& +...
\label{xa1ratemnk}
\end{eqnarray}
One thus obtains the standard large deviation form with respect to the volume $L_n^d=(AB)^n$ of Eq. \ref{largedevm}
and the corresponding rate function $I^-(x) $ displays the tail behavior
\begin{eqnarray}
I^-(x) \opsimeq_{ x \to -\infty }
&& \lambda_0^-   \vert x \vert -  \left( \nu^-_0 - \frac{B-1}{AB-1}   \right)  \ln \vert x \vert
- \ln(K_{0}^-) + \frac{B-1}{AB-1} \ln(\lambda_0^- )
\nonumber \\ &&
   +  \frac{B}{2(AB-1)} \ln A
-\frac{B(A-1)}{AB-1} \left( 
\frac{\ln (AB)}{2 (AB-1)} 
+ \ln \left(\sqrt{ \frac{ 2 \pi } {\nu^-_0 - \frac{B-1}{AB-1}  } } \right)
\right)
\label{xa1ratem}
\end{eqnarray}
instead of Eq. \ref{tailratem}.


\section{ Conclusions }

\label{sec_conclusion}

In this paper, we have revisited the renormalization rule for the ground-state energy of the Directed Polymer model 
on a hierarchical lattice of parameters $(A,B)$ in order to analyze the renormalization flows for the tails of probability distributions 
as a function of the initial condition at generation $n=0$.
In each case, the explicit solution has allowed to extract 
the scalings involved in the large deviations properties 
and the tail behaviors of the corresponding rate functions. 
Our main conclusions can be summarized as follows :

(i) the generic large deviation form with respect to the length $L_n$ for the tail $x \to +\infty$ 
emerges only for $\alpha^+ \geq 1$, while the stretched exponential 
$0< \alpha^+ <1$ initial conditions lead to anomalous large deviations in $L_n^{\alpha^+} $.

(ii) the generic large deviation form with respect to the volume $L_n^d$ for the tail $x \to -\infty$
emerges only for $\alpha^- \geq 1$, while the stretched exponential 
$0< \alpha^- <1$ initial conditions lead to anomalous large deviations in $L_n^{d-1+\alpha^-} $.

This example shows that it is interesting to analyze the renormalization flows of disordered systems at the level of large deviations, in order to go beyond the region of typical fluctuations that have been much studied in the past.


\appendix

\section{ Tail analysis for the empirical average of a finite number $A$ of random variables }

\label{app_convol}

In this Appendix, we consider a finite number $A$ of independent random variables $x_a$
distributed with some probability distribution ${\cal P}(x) $ whose tail for $x \to +\infty$ is characterized by the function $f(x)$
\begin{eqnarray}
{\cal P}(x)  \opsimeq_{ x \to + \infty } e^{- f(x) }
\label{tailf}
\end{eqnarray}
The empirical average
\begin{eqnarray}
x \equiv \frac{1}{A} \sum_{a=1}^A x_a
\label{xempi}
\end{eqnarray}
is distributed with the convolution
\begin{eqnarray}
 {\cal A}(x)= \int_{-\infty}^{+\infty} dx_1 ... \int_{-\infty}^{+\infty} dx_A 
{\cal P}(x_1) ... {\cal P}(x_A) \delta\left( x- \frac{1}{A}\sum_{a=1}^A x_a   \right)
\label{Aconvolution}
\end{eqnarray}
The tail behavior as $x \to +\infty$ of this convolution depends on the tail of Eq. \ref{tailf}. 
For concreteness, it will be convenient to consider the family
\begin{eqnarray}
{\cal P} (x) \opsimeq_{x \to +\infty} K x^{\nu-1} e^{- \lambda x^{\alpha} } 
\label{expalpha}
\end{eqnarray}
 so that the corresponding tail
function $f(x)$ of Eq. \ref{tailf}
and its second derivative read
\begin{eqnarray}
f(x) && =  \lambda x^{\alpha} +(1-\nu) \ln x - \ln K
\nonumber \\
f''(x) &&= \lambda \alpha (\alpha-1) x^{\alpha-2} +\frac{\nu-1}{x^2}
\label{fexpalpha}
\end{eqnarray}

\subsection{ The 'democratic' formula for $\alpha >1$ }

The 'democratic' formula obtained in Ref \cite{frisch} 
\begin{eqnarray}
 {\cal A}^{democratic}(x)  \opsimeq_{x \to + \infty}  e^{ -A f(x) } \sqrt{A} \left( \frac{2 \pi} { f''(x) }  \right)^{\frac{A-1}{2} }
\label{frischa}
\end{eqnarray}
can be understood from two points of view.

\subsubsection{ 'Democratic' saddle-point analysis of Ref \cite{frisch} }

The formula of Eq. \ref{frischa} has been derived in Ref \cite{frisch} 
from the saddle-point evaluation of the convolution of Eq. \ref{Aconvolution}
around the symmetric solution $x_a=x$ for $a=1,2..,A$ with the 
two validity conditions (see \cite{frisch} for very detailed discussions and various formulations of the validity conditions)
\begin{eqnarray}
 && f''(x)  >  0
\nonumber \\
 && x^2 f''(x)  \opsimeq_{x \to +\infty}  + \infty
\label{convex}
\end{eqnarray}
For the special family of Eq. \ref{expalpha},
these conditions are satisfied only in the region
\begin{eqnarray}
\alpha>1
\label{alphabigger}
\end{eqnarray}
while they are not satisfied for $0<\alpha \leq 1$.

\subsubsection{ Alternative derivation via the tail $k \to +\infty$ of the cumulant generating function }

Another way to understand Eq. \ref{frischa}
involves the cumulant generating function $\phi(k)$  
\begin{eqnarray}
e^{\phi(k)} \equiv \int_{-\infty}^{+\infty} dx e^{k x } {\cal P}(x)=\int_{-\infty}^{+\infty} dx e^{k x -f(x) }
\label{scaledphik}
\end{eqnarray}
For $\alpha>1$, this cumulant generating function exists even for large $k$,
and the tail for $k \to + \infty$ is determined by the tail for $x \to + \infty$ of Eq. \ref{tailf}
via the saddle-point evaluation of Eq. \ref{scaledphik}
around the large saddle-point value $x_k$ satisfying
\begin{eqnarray}
f'(x_k) && = k
\label{scaledphikcol}
\end{eqnarray}
that leads to the asymptotic result
\begin{eqnarray}
e^{\phi(k)} \opsimeq_{k \to + \infty} 
\int_{-\infty}^{+\infty} dx e^{k x_k -f(x_k) - \frac{(x-x_k)^2}{2} f''(x_k) }
= e^{k x_k -f(x_k) } \sqrt{ \frac{2 \pi} {f''(x_k)   } }
\label{scaledphikcolres}
\end{eqnarray}

The scaled cumulant generating function associated to the empirical average of Eq. \ref{Aconvolution}
is simply given by the power $A$ of Eq. \ref{scaledphik}
\begin{eqnarray}
 \int_{-\infty}^{+\infty} dx e^{A k  x  } {\cal A}(x)= \left( \int_{-\infty}^{+\infty} dx e^{k x } {\cal P}(x) \right)^A =e^{A \phi(k) } 
\label{scaledphiv}
\end{eqnarray}
Eq \ref{scaledphikcolres} then yields that the tail for $k \to + \infty$ is given by
\begin{eqnarray}
 \int_{-\infty}^{+\infty} dx e^{k A x  } {\cal A}_n(x)
\opsimeq_{k \to + \infty} \left(  e^{k x_k -f(x_k) } \sqrt{ \frac{2 \pi} {f''(x_k)   } }  \right)^A
 =e^{A k x_k - A f(x_k) } \left(  \frac{2 \pi} {f''(x_k)  }   \right)^{\frac{A-1}{2} }
\sqrt{ \frac{2 \pi} {f''(x_k) }} 
\label{scaledphivtail}
\end{eqnarray}
that corresponds indeed to the saddle-point evaluation of the tail of Eq. \ref{frischa}.

\subsection{ The 'monocratic' formula for $0 < \alpha <1$ }

The 'monocratic' formula corresponds to the cases where the tail $x \to +\infty$ of the convolution of Eq. \ref{Aconvolution}
is dominated by the drawing of the anomalously large value $y \simeq Ax$ for the maximum
of the $A$ variables $(x_1,..,x_A)$, while the other $(A-1)$ values remain typical,
so that one obtains the tail behavior
\begin{eqnarray}
 {\cal A}^{monocratic}(x)  \opsimeq_{x \to + \infty}  A \int dy  {\cal P}(y) \delta\left( x- \frac{y}{A}   \right)
= A^2 {\cal P}(Ax) = A^2e^{ - f(Ax) } 
\label{monocratic}
\end{eqnarray}
that indeed gives a bigger result than the 'democratic' formula of Eq. \ref{frischa}  for $0 < \alpha<1$.

\subsection{ The intermediate case $ \alpha =1$ when $\nu>0$ }

For the intermediate case $\alpha=1$ of Eq. \ref{expalpha}
\begin{eqnarray}
 {\cal P}(x) \opsimeq_{x \to + \infty} K x ^{\nu-1} e^{- \lambda x }
\label{xalpha1}
\end{eqnarray}
one can use neither the 'democratic' formula nor the 'monocratic' formula described above.
For $\nu>0$, the cumulant generating function $\phi(k)$ of Eq. \ref{scaledphik} 
exists only for $k<\lambda$ and diverges as $k \to \lambda$.
This singularity as $ k \to \lambda$ 
is then governed by the tail $x \to +\infty$ of Eq. \ref{xnalpha1}
that one assumes to be valid in the region $x>C$ (where $C$ is some fixed large constant $C>0$)
\begin{eqnarray}
e^{\phi(k)} && \opsimeq_{k \to \lambda} \int_C^{+\infty} dx K x ^{\nu-1} e^{- (\lambda-k)  x }
= \frac{K}{ (\lambda-k)^{\nu} } \int_{C (\lambda-k)  } ^{+\infty} dt  t ^{\nu-1} e^{- t }
 \opsimeq_{k \to \lambda}  \frac{K \ \Gamma(\nu) }{ (\lambda-k)^{\nu}}
\label{scaledphising}
\end{eqnarray}

The scaled cumulant generating function of Eq. \ref{scaledphiv}
associated to the empirical average of Eq. \ref{Aconvolution}
then displays the singularity
\begin{eqnarray}
 \int_{-\infty}^{+\infty} dx e^{A k  x  } {\cal A}(x) = e^{A \phi(k) } 
\opsimeq_{k \to \lambda}  \frac{ \left[ K \  \Gamma(\nu)\right]^A }{ (\lambda-k)^{A\nu}}
\label{scaledphiva1}
\end{eqnarray}
that corresponds to the following tail as $x \to +\infty$
\begin{eqnarray}
 {\cal A}(x)\opsimeq_{x \to + \infty} \frac{ A^{A \nu } 
\left[ K \Gamma(\nu)\right]^A }{\Gamma(A \nu)} x^{A \nu-1}e^{- A \lambda x }
\label{ana1}
\end{eqnarray}

\subsection{ Final remark on the similarities and differences with the large deviations of the empirical average }

In this Appendix, we have considered as in Ref  \cite{frisch}
the problem of the tail $x \to +\infty$ of the empirical average of a finite number $A$ of independent variables,
while the standard large deviations problem for the empirical average focuses instead 
on a large number $A \to +\infty$ of independent variables, while $x$ remains finite.
The two problems are thus clearly different, but they nevertheless 
display some similarities as discussed in detail in Ref \cite{frisch},
and the two democratic/monocratic behaviors 
have also been much studied in the large deviation regime \cite{nagaev,evans2008,nina,godreche,evans2014,c_largedevasym}.

\end{document}